\documentclass[prl,twocolumn,showpacs]{revtex4}

\usepackage{amssymb,amsmath,bm,epsfig,graphics,subfigure}

\begin{document}

\title{Casimir Effect for Massless Fermions in One Dimension: A Force Operator Approach}
\author{Dina Zhabinskaya}
    \email{dinaz@physics.upenn.edu}
 \author{Jesse M. Kinder and E.J. Mele}   
    \affiliation{Department of Physics and Astronomy \\ University of Pennsylvania, Philadelphia PA 19104}
\date{\today}

\begin{abstract}
 We calculate the Casimir interaction between two short range scatterers embedded in a background of one dimensional
  massless Dirac fermions using a force operator approach.  We obtain
   the force between two
finite width square barriers, and take the limit of zero width and infinite
potential strength to study the Casimir force mediated by the fermions.
For the case of identical
scatterers we recover the conventional attractive one dimensional
Casimir force. For the general problem with inequivalent
scatterers we find that the magnitude and sign of this force
depend on the relative spinor polarizations of the two scattering
potentials which can be tuned to give an attractive, a repulsive, or a
compensated null Casimir interaction.
\end{abstract}

\maketitle

Boundaries modify the spectrum of zero point fluctuations of a
quantum field, resulting in fluctuation-induced forces and
pressures on the boundaries that are known generally as Casimir
effects [\onlinecite{casimir.book}].  When sharp boundaries conditions are used to model the Casimir effect, they yield perfect reflection of the incident
propagating quantum field at all energies [\onlinecite{casimir.book}].
However, in many physical applications this hard-wall limit is not
appropriate; of special interest in the present work are
interactions between localized scatterers in one dimension that
have energy-dependent scattering properties controlled by the
strength, range  and shape of the potential. Along this line,
previous work has recognized that the finite reflectance of
partially transmitting mirrors provides a natural high energy
regularization scheme for computing the effect of sharp reflecting
boundaries on the zero point energy of the electromagnetic field
[\onlinecite{Reynaud,delta}]. In more recent work, Sundberg and Jaffe
approached the problem of computing the effect of confining
boundary conditions on a degenerate gas of fermions in one
dimension as the limiting behavior for  rectangular barriers of
finite width and height. Interestingly, they encounter a
divergence of the Casimir energy in the zero width limit (a sharp
boundary) even for finite potential strength [\onlinecite{Jaffe}].

In this Rapid Communication we address the problem of Casimir interactions
between scatterers mediated  by a one-dimensional Fermi gas. The
fermions in our calculation are massless Dirac fermions appropriate
to describe, for example, the (single-valley) electronic spectrum of
a metallic carbon nanotube.  We employ the Hellmann-Feynman theorem to
calculate the force, rather than energy,
 of interaction between two scatterers
as a function of their separation $d$. This approach renders our
calculation free from ultraviolet divergences even for the
limiting case of sharp scatterers.  We demonstrate that for the
case of identical scatterers, this formalism recovers the
well known attractive $1/d^2$ Casimir force in one dimension.
Furthermore, we find that for Dirac fermions the internal
structure of the matrix-valued scattering potential admits a long
range Casimir interaction which can also be repulsive or even
compensated. This provides a physical situation where the Casimir
interaction is continuously tunable from attractive to repulsive
by variation of an internal control parameter, realizing the known
bounds for the one dimensional Casimir interaction as two limiting
cases. The results may be relevant for indirect interactions
between defects and adsorbed species on carbon nanotubes.

The fermions in our model are massless one-dimensional Dirac
fermions described by the Hamiltonian
\begin{equation}
\Big( -i\sigma_x\partial_{x}+ \hat V(x)-E\Big)\Psi_k(x)=0,
\label{eq:H}
\end{equation}
where we set $\hbar=c=1$.
In graphene and carbon nanotubes the spinor polarizations 
describe the internal degrees of freedom generated by the two-sublattice structure in its primitive cell. When $\hat V(x)=0$, the eigenstates of $\mathcal{H}_o$ are plane waves
multiplying two-dimensional spinors, $\Psi_k(x)=\Phi_k
e^{ikx}/\sqrt{2\pi}$.  When the chemical potential is fixed at $\mu =0$, the filled Dirac sea has $E=-|k|$ with
$\Phi^{T}_{\pm k}=(1, \mp 1)/\sqrt{2}$.

The general form of the potential entering (1) is $\hat V(x) =
V_{o}(x)\hat{\texttt{I}} +\vec{V}(x) \cdot\vec{\sigma}$. The
$\sigma_x$ part of the potential can eliminated by a gauge
transformation [\onlinecite{Jaffe}], and a scalar potential
proportional to the identity matrix produces no backscattering in
the massless Dirac equation.  Therefore, we consider potentials
for which $\vec V$ lies in the $yz$-plane. In this paper, we consider
the effects of the orientation of the potential determined by
angle $\phi$. Thus, a square barrier potential located between
points $x_1$ and $x_2$ is written as
\begin{equation}
\hat{V}(x,\phi)=\hat{V}(\phi)\theta(x-x_1)\theta (x_2-x),  \label{eq:H1}
\end{equation}
where $\hat{V}(\phi)=Ve^{i\sigma_x\phi /2}\sigma_ze^{-i\sigma_x\phi
/2}$, and $\theta(x)$ is a step function.

To study the force on a square well scatterer we use the
Hellmann-Feynman theorem,
$\langle\partial\hat{\mathcal{H}}(\lambda)/\partial\lambda\rangle=\partial
E/\partial\lambda$ [\onlinecite{HF}]. Taking the control parameter
$\lambda = (x_1 +x_2)/2 = \bar x$, the ground state average gives
the force acting on a rigid barrier. For a barrier with sharp
walls the expectation value becomes
\begin{eqnarray}
\left\langle\Psi(x)\right|\frac{\partial\mathcal{\hat{H}}}{\partial \bar x}\left|\Psi(x)\right\rangle=\left\langle\Psi(\bar x +a/2)\right|\hat{V}\left|\Psi(\bar x +a/2)\right\rangle \nonumber \\
-\left\langle\Psi(\bar x -a/2)\right|\hat{V}\left|\Psi(\bar x
-a/2)\right\rangle, \label{eq:gen.force}
\end{eqnarray}
where $\hat{V}$ is the square barrier potential, $\bar x$ is its
center and $a$ is its width. The total force is the expectation
value of this force operator,
$\hat{F}=-\partial{}\hat{\mathcal{H}}/\partial{} \bar x$, summed
over all the occupied states;
  Eq.~(\ref{eq:gen.force}) then gives the difference between the pressures exerted on the right
  and the left sides of the barrier. For potentials of general shape a similar expression can be
  developed in terms of an integral over the scattering region.

First, we apply   Eq.~(\ref{eq:gen.force}) to calculate the force
on an isolated barrier.  The eigenstates
are represented as linear combinations of right and left moving
solutions of $\mathcal{H}_o$:
$\Psi(x)=\frac{1}{\sqrt{2\pi}}(\alpha_k\Phi_k e^{ikx}+\beta_k\Phi_{-k}
e^{-ikx})$, where $\alpha_k$ and $\beta_k$ represent the
amplitudes of the counterpropagating waves in each region. The
$yz$ polarized potential defined in Eq.~(\ref{eq:H1}) gives
$\hat{V}(\phi)\Phi_{\pm k}=Ve^{\pm i\phi}\Phi_{\mp k}$, so the
general expression for the expectation values in
Eq.~(\ref{eq:gen.force}) at some position $x$ is
\begin{equation}
\langle\Psi(x)|\hat{V}(\phi)|\Psi(x)\rangle=\frac{V}{\pi}\mathcal{R}e[\alpha_k\beta_k^*e^{i(2kx+\phi)}].  \label{eq:expectation}
\end{equation}

We use a transfer matrix to obtain the coefficients $\alpha_k$ and
$\beta_k$ entering Eq.~(\ref{eq:expectation}). The transfer matrix
is defined so that $\Psi(x_2)=T\Psi(x_1)$, where $x_1$ and $x_2$ are the
left and right boundaries of a barrier, respectively
; $T$
is calculated by integrating Eq.~\ref{eq:H},
\begin{equation}
T=P_x\exp\left(i\int_{x_1}^{x_2}dx\sigma_{x}[E-\hat V(x)]\right),
\end{equation}
where $P_x$ is a spatial ordering operator.  For the square
potential of width $a$ defined in Eq.~(\ref{eq:H1}), the transfer
matrix for negative energy states is
\begin{equation}
T=\cos(qa)-\frac{i\sigma_x k-\vec{\sigma}\cdot(\hat{x}\times\vec{V})}{q}\sin(qa),
\end{equation}
where $\vec{V}=V(0,\sin\phi,\cos\phi)$ defines a potential in the
$yz$-plane, $q=\sqrt{k^2-V^2}$,  and $k>0$.

From the transfer matrix we calculate the scattering matrix $S$,
which gives the transmitted ($t$) and reflected ($r$) amplitudes
for wave incident on the barrier from the right and from the left.
The unitary S-matrix for a single square
barrier is
\begin{equation}
S_1=
\begin{pmatrix}
te^{-ika} & re^{-i(2kx_2+\phi)} \\ re^{i(2kx_1+\phi)} & te^{-ika} \\
\end{pmatrix}
.
\label{eq:S-matrix}
\end{equation}
The transmission and reflection coefficients can then be
parameterized $t=\tau e^{i\eta}$ and
$r=i\sqrt{1-\tau^2}e^{i\eta}$, where
\begin{equation}
\tau =\frac{\lambda}{(V^2\cosh^2\lambda a-k^2)^{1/2}},\eta =\tan^{-1}\Big(\frac{k\tanh\lambda a}{\lambda}\Big),
\end{equation}
with $\lambda=-iq=\sqrt{V^2-k^2}$.  To obtain the hard-wall limit,
 we fix $\Gamma=V a$, and take $\Gamma\to\infty$.
In this limit, $|r|^2\to 1$ and $|t|^2\to 0$ at all energies.

\begin{figure}[h]
\includegraphics*[width=2.8in,height=1.8in]{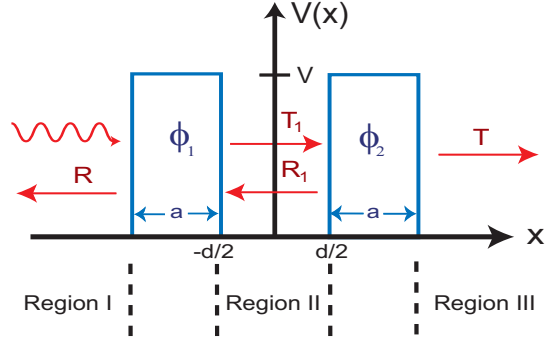}
\caption{Scattering of massless Dirac fermions (incoming from the
left) between two square barriers of height $V$, width $a$, and
separation $d$.  The two potentials defined in Eq.~(\ref{eq:H1})
have a spinor polarization determined by angle $\phi$.  The
reflection and transmission coefficients are labeled in each
scattering region.} \label{fig:potentials}
\end{figure}

For a single barrier, 
the contributions to the force from the
particles incoming from the right and the left cancel,
resulting in no net force. A nonzero force
arises from the multiple reflection of electron waves between two
barriers. An illustration of a scattering process for two
square potentials with different spinor polarizations $\phi_1$ and
$\phi_2$ separated by distance $d$  is shown in
Fig.~\ref{fig:potentials}. The contributions from waves incoming from the right are also included in the calculation.

The S-matrix for the two-barrier system [\onlinecite{Datta}] in Fig.~\ref{fig:potentials} is
\begin{equation}
S_2=
\begin{pmatrix}
T & Re^{i\phi_1} \\ Re^{-i\phi_2} & T \\
\end{pmatrix}
.
\label{eq:S2}
\end{equation}
The total reflection and transmission coefficients shown in regions I and III of Fig.~\ref{fig:potentials} are given by
\begin{equation}
T=\frac{t^2}{1-r^2e^{i\nu}},~R=re^{-ik(2a+d)}\Big(1+\frac{t^2e^{i(2ka+\nu)}}{1-r^2e^{i\nu}}\Big),
\end{equation}
where $\nu=2kd+\delta\phi$ and $\delta\phi\equiv\phi_2-\phi_1$.
$T_1$ and $R_1$ in region II of Fig.~\ref{fig:potentials} are given by
\begin{equation}
T_1=\frac{t}{1-r^2e^{i\nu}},~~R_1=\frac{rte^{i(kd+\phi_2)}}{1-r^2e^{i\nu}}.
\end{equation}
The coefficients for the waves incoming from the left ($R_1$ and $T_1$), and the ones incoming from the right ($R_1'$ and $T_1'$) are related by
$R_1'=R_1e^{-i(\phi_1+\phi_2)}$ and $T_1'=T_1$.

To calculate the force in the two-barrier problem we fix the position of the
left barrier in Fig.~\ref{fig:potentials} and differentiate the
Hamiltonian with respect to $d$. To obtain the total force, we sum over the occupied states of the filled Dirac sea at fixed chemical potential. 
We find that the force between two
square barriers of finite height and width is
\begin{eqnarray}
F=&-2V&\int_{0}^{\infty}\frac{dk}{2\pi}\mathcal{R}e[Re^{ik(d+2a)}\nonumber \\
& & -R_1T^{*}_1e^{-i(kd+\phi_2)}(1+e^{i\nu})]. \label{eq:forceR}
\end{eqnarray}
The first term in the integrand arises from the
exterior modes pushing the two barriers together. The second term
accounts for the confined modes in between the barriers pushing
them apart.  Since incoming waves are fully transmitted at high
energies for barriers of finite height and width,
 the integral in Eq.~(\ref{eq:forceR}) converges even in the case
of sharp barriers ($a\to 0$), with $\Gamma=Va$ fixed.  Thus, the
reflection coefficient  provides a natural cutoff for the
computation of the force (though not the energy [\onlinecite{Jaffe}])
even in the limit of infinitely high
barriers.

The Casimir force for hard-wall boundary conditions requires the
limits of infinite barrier strength $\Gamma\to\infty$ and zero width
$a\to 0$. This limit enforces a vanishing current at the boundaries, the so-called bag boundary conditions.
 Since the force in Eq.~(\ref{eq:forceR}) is multiplied by $V$, we keep
terms to $\mathcal{O}(k/V)$ in the integrand. The first term in
Eq.~(\ref{eq:forceR}) becomes proportional to $k$, thus implying a
continuous spectrum of modes scattering off the barriers from the
outside.  The second term exhibits resonances that arise from the
quantized modes between the boundaries.  These resonances, similar
to ones seen in Fabry-Perot cavities, are represented by
Dirac delta functions [\onlinecite{delta}] to
constrain the $k$ integration
\begin{equation}
\lim_{\tau\rightarrow 0}\frac{\tau^2}{|1+(1-\tau^2)e^{i(\nu+2\eta)}|^2}=\frac{\pi}{2d}\sum_{n=0}^{\infty}\delta(k-k_n),
\label{eq:resonance}
\end{equation}
where $k_n=\pi[n+(1-\delta\phi/\pi)/2]/d$, and $\eta\to 0$
in the limit of infinite potential strength. Here $\delta \phi$ is
the difference in the spinor polarizations of the two scattering
potentials, and $\delta \phi = 2 \pi n$ denotes the situation for
identical scatterers. An incoming wave vector satisfying
the resonance condition in Eq.~(\ref{eq:resonance}) gets fully
transmitted through the two-barrier system.  The modes in between
the barriers, on the other hand, are fully reflected yielding the
appropriate quantization condition. Combining these results we
obtain
\begin{equation}
F=2\int_0^{\infty}\frac{dk}{2\pi}k\Big[1-\frac{\pi}{d}\sum_{n=0}^{\infty}\delta(k-k_n)\Big]+\mathcal{O}\Big(\frac{1}{V}\Big). \label{eq:force.quant}
\end{equation}

\begin{figure}[h]
\includegraphics*[width=2.6in,height=1.9in]{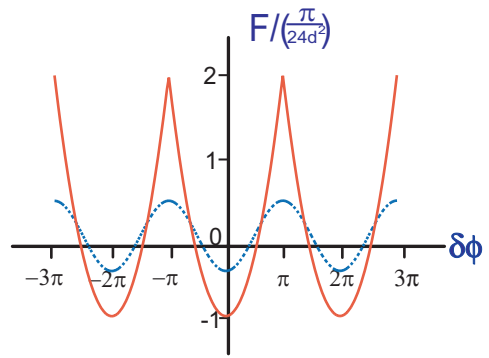}
\caption{Force between two barriers as a function of their relative spinor polarization $\delta\phi$.  The solid and dashed lines represent the forces in Eq.~(\ref{eq:final.force}) and Eq.~(\ref{eq:smallV}), respectively.  The magnitude of the force in the $\Gamma\ll 1$ limit, the dashed curve, is rescaled to $\Gamma=1/2$ so the two curves can be compared.}
\label{fig:force.plot}
\end{figure}

The Casimir force in Eq.~(\ref{eq:force.quant}) can be calculated
by applying the generalized Abel-Plana formula,
\begin{eqnarray}
\int_{0}^{\infty}tdt-\sum_{n=0}^{\infty}(n+\beta)= ~~~~~~~~~~~~~~~~~~~\nonumber \\
-\int_{0}^{\infty}tdt\Big(\frac{\sinh(2\pi t)}{\cos(2\pi\beta)-\cosh(2\pi t)}+1\Big), \label{eq:Abel}
\end{eqnarray}
which is valid for $0\leq\beta<1$.  Due to the rapid convergence
of the integral in Eq.~(\ref{eq:Abel}), the result does not require an
introduction of an explicit ultraviolet cutoff function
[\onlinecite{Abel}].  More generally, since the reflection
coefficient vanishes at high energy it will regularize  the
calculation of the force. Using Eq.~(\ref{eq:Abel}) we obtain the
force for two barriers satisfying bag boundary conditions,
\begin{equation}
F=-\frac{\pi}{24d^2}\Big[1-3\Big(\frac{\delta\phi}{\pi}\Big)^2\Big] \label{eq:final.force}
\end{equation}
for $-\pi\leq\delta\phi <\pi$ beyond which it is periodic. We also explore the force between two scatterers of finite height and width.  In the small barrier strength limit the force becomes
\begin{equation}
F=-\frac{\Gamma^2\cos(\delta\phi)}{2\pi d^2}\Big[1+\mathcal{O}\Big(\frac{a}{d}\Big)\Big]. \label{eq:smallV}
\end{equation}
The force in the limits of $\Gamma\to\infty$ and $\Gamma\ll 1$ for $a\to 0$ is plotted for three periods in $\delta\phi$ in Fig.~\ref{fig:force.plot}.

The scaling of the
force with distance as $1/d^2$ and the ratio of $1/2$ between the
repulsive and attractive forces are universal results for massless
 one-dimensional fluctuating fields in the limit $d\gg a$. 
When the range of 
the potentials becomes comparable to their separation, the first order correction 
due to the shape of the scatterer scales with $\delta F/F\sim a/d$ 
as seen in  Eq.~(\ref{eq:smallV}), analogous to a multipole expansion of 
an electrostatic interaction.  

The relative orientation can be expressed as
$\delta\phi=\cos^{-1}(\vec{V_1}\cdot\vec{V_2}/(|\vec{V_1}| \cdot
|\vec{V_2}|))$. When the two potentials are aligned at $\delta\phi
=2\pi n$ we have $F=-\pi/24d^{2}$. This yields the attractive
fermionic Casimir force as found in Ref. [\onlinecite{Jaffe}].
When $\delta \phi = (2n+1) \pi$ the relative polarization of the defect
potentials is antiparallel and $F=\pi/12d^2$, i.e. a repulsive Casimir
force is obtained.  An analog of our result for a
one-dimensional bosonic field is obtained by imposing mixed
Dirichlet and Neumann boundary conditions where attractive and
repulsive Casimir forces are found for like and unlike boundary
conditions, respectively [\onlinecite{boyer2}]. 
 A Casimir force that oscillates as a function
of defect separation $d$ is known to arise from large momentum
backscattering (Friedel oscillations) of the Fermi gas
[\onlinecite{Recati}]. However, the interaction we calculate here
is  monotonic as a function of distance.
In our calculation, the  magnitude and sign of the force varies as a function of the relative polarization of two scatters at a fixed distance.  As shown in Fig.~\ref{fig:force.plot} this behavior occurs for both finite barriers and hard-wall boundaries.

\begin{figure}[h]
\includegraphics*[width=2.4in,height=2in]{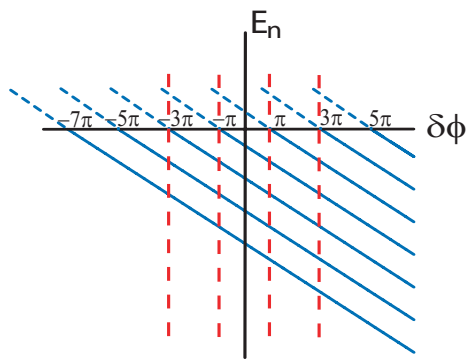}
\caption{Quantized energy bands for massless Dirac fermions due to
hard-wall boundary conditions as a function of the relative
polarization of the two potentials $\delta\phi$.  Solid lines
denote the energy levels of the filled Dirac sea.  Vertical dashed
lines define $2\pi$ periodic states where the number of
occupied states changes by one.} \label{fig:energy.bands}
\end{figure}

The cusps seen in Fig.~\ref{fig:force.plot} at the odd multiples of $n$
result from a sum over the discrete number of  energy levels $E_n(\delta\phi)$ .  The
energy bands found in Eq.~(\ref{eq:resonance}) cross zero energy
at $\delta\phi=(2n+1)\pi$ as shown in Fig.~\ref{fig:energy.bands}.
At fixed chemical potential, with negative energy states of the
Dirac sea occupied, the number of states changes by one in each
$2\pi$ periodic region indicated by dotted vertical lines in
Fig.~\ref{fig:energy.bands}. Consequently, the force exhibits a
discontinuity in slope in Fig.~\ref{fig:force.plot} exactly at the
values of $\phi$ at which there is a jump in the number of
occupied energy levels. When the barrier strength is finite, the cusps in the force disappear.
The resonance condition resulting in quantized states 
between the barriers is only valid for hard-wall boundaries.  
Note, the energy states between finite barriers exhibit a continuous spectrum.

The interaction Eq.~(\ref{eq:final.force}) is likely to be
important for defect interactions on carbon nanotubes, and
possibly for other one-dimensional systems as well. Reinserting
dimensional factors this force corresponds to an interaction
energy $E_c=-\pi\hbar v_F/24d$ for two identical scatterers. With
$\hbar v_F \sim 5.4 \, {\rm eV \cdot \AA}$ this gives an energy of
$1.4 \, {\rm meV}$ at a range $d=50 \, {\rm nm}$. Note that its
spatial form  follows the same scaling law as the Coulomb
interaction between uncompensated charges, but it is reduced by a
factor $\pi \hbar v_F/24 e^2 \sim .05$. Thus, for charge neutral
dipoles $p=es$ whose electrostatic interactions scale as $E_{\rm
d} \sim -p^2/d^3 = -(e^2/d) \times (s/d)^2$, they are dominated by
the Casimir interaction in the far field $d \gtrsim 5s$.
Similarly, this one-dimensional Casimir interaction completely
dominates the familiar van der Waals interactions between charge
neutral species that are mediated by the fluctuations of the
exterior three dimensional electromagnetic fields.

In order to fully understand the Casimir effect between defects on
carbon nanotubes, one needs to consider the symmetry and range of
the potentials produced by localized defects.  The spinor
polarization discussed in this paper is determined by the form
of the impurity potential: $\sigma_z$ and $\sigma_y$
potentials define a sublattice-asymmetric and bond-centered
defects, respectively. In addition, the electronic spectrum
contains two distinct Fermi points at inequivalent corners of the
two dimensional Brillouin zone. Short-range potentials couple the
two Fermi points resulting in intervalley scattering
[\onlinecite{Ando}]. Therefore, both the structure of the defects
and the effect of intervalley scattering determine the sign and
magnitude of the Casimir interaction. In the context of our model,
a sharp potential is one with a range on the order of the tube
radius $R$ for which the effects of intervalley scattering are
suppressed by a factor of $a_c/R$, where $a_c$ is the width of the
graphene primitive cell. Atomically sharp scatterers, on the other hand,
 will usually
require a treatment of the effects of intervalley as well as
intravalley scattering.

To summarize, we introduced a  force operator approach for
calculating the Casimir effect and obtained the
fluctuation-induced force between two finite square barriers
mediated by massless Dirac fermions in one dimension. In taking
the limit of sharp barriers of infinite strength we obtained a Casimir
force that scales as $1/d^2$, and is tunable from attractive to repulsive
form as a function of the relative spinor polarizations of the two
scattering potentials.

This work was supported by the Department of Energy under grant
DE-FG02-ER45118.

\end{document}